\documentstyle{mn}

\input epsf

\begin{document}
\journal{Sussex preprint SUSSEX-AST 95/6-2, astro-ph/9506091}
\title[Open cold dark matter models]{Open cold dark matter models}
\author[A. R. Liddle et al.]{Andrew R.~Liddle,$^1$ David H.~Lyth,$^2$ David 
Roberts$^2$ and Pedro T.~P.~Viana$^1$\\
$^1$Astronomy Centre, University of Sussex, Falmer, Brighton 
BN1 9QH\\
$^2$School of Physics and Materials, University of Lancaster, Lancaster LA1 
4YB}
\maketitle
\begin{abstract}
Motivated by recent developments in inflationary cosmology indicating the 
possibility of obtaining genuinely open universes in some models, we compare 
the predictions of cold dark matter (CDM) models in open universes with a 
variety of observational information. The spectrum of the primordial 
curvature perturbation is taken to be scale invariant (spectral index $n=1$), 
corresponding to a flat inflationary potential. We allow arbitrary variation 
of the density parameter $\Omega_0$ and the Hubble parameter $h$, and take 
full account of the baryon content assuming standard nucleosynthesis. We 
normalize the power spectrum using the recent analysis of the two year {\it 
COBE} DMR data by G\'{o}rski et al. We then consider a variety of 
observations, namely the galaxy correlation function, bulk flows, the 
abundance of galaxy clusters and the abundance of damped Lyman alpha systems. 
For the last two of these, we provide a new treatment appropriate to open 
universes. We find that, if one allows an arbitrary $h$, then a good fit is 
available for any $\Omega_0$ greater than 0.35, though for $\Omega_0$ close 
to 1 the required $h$ is alarmingly low. Models with $\Omega_0 < 0.35$ seem 
unable to fit observations while keeping the universe over $10$ Gyr old; this 
limit is somewhat higher than that appearing in the literature thus far. If 
one assumes a value of $h > 0.6$, as favoured by recent measurements, 
concordance with the data is only possible for the narrow range $0.35 < 
\Omega_0 < 0.55$. We have also investigated $n \neq 1$; the extra freedom 
naturally widens the allowed parameter region. Assuming a range $0.9<n<1.1$, 
the allowed range of $\Omega_0$ assuming $h > 0.6$ is at most $0.30 < 
\Omega_0 < 0.60$. 
\end{abstract}

\section{Introduction}

Even before the announcement of the detection of microwave background
anisotropies by the DMR experiment on the {\it COBE} satellite 
\cite{COBE}, it was realized that structure formation models based on cold 
dark matter (CDM) and a flat spectrum of primordial perturbations fared 
considerably better against the data if the matter density was reduced by a 
factor of around three. Most studies of this possibility invoked a 
cosmological constant to restore spatial flatness \cite{ESM,KGB}, with little 
attention being directed to the possibility that the cosmological constant 
may be redundant and the low-density model implemented in a genuinely open 
universe. This produces the same shape of perturbation spectrum on scales 
well below the curvature radius, but a different normalization and redshift 
dependence. 

The reluctance to study such models (though general arguments in favour of an 
open universe were developed, e.g. Coles \& Ellis \shortcite{ce} and Primack 
\shortcite{Prim}) arose from a widespread belief that inflation, the most 
plausible candidate for generating the initial density perturbations, could 
give rise to an open universe only in exceptionally fine-tuned circumstances. 
However, open universe inflation models have received renewed interest 
recently, and in particular attention has been drawn (Sasaki et al. 1993a, 
1993b; Tanaka \& Sasaki 1994; Bucher, Goldhaber \& Turok 1995; Yamamoto, 
Sasaki \& Tanaka 1995a; Sasaki, Tanaka \& Yamamoto 1995; Yamamoto, Tanaka \& 
Sasaki 1995b; Bucher \& Turok 1995) to the bubble nucleation model 
\cite{bubblea,bubbleb,bubblec,bubbled,bubblee,bubblef}. In contrast with the 
situation for ordinary models of inflation \cite{lyst,RPb}, this model 
predicts the present value of the density parameter in terms of the scalar 
field potential, without any reference to initial conditions. The price that 
one pays for this is a non-generic scalar field potential, which will be even 
more difficult than usual to realize in the context of a sensible particle 
physics model.

In order to compare structure formation models based on open universe 
inflation models with observational data, it is crucial to be able to 
normalize the amplitude of the power spectrum to the {\it COBE} observations 
of cosmic microwave background (CMB) fluctuations \cite{B94}, which are by 
far the most accurate available. Anisotropy calculations in an open universe 
present many technical difficulties and progress to the result has 
consequently been slow. However, an accurate normalization is now available 
through the work of G\'{o}rski et al. (1995, henceforth GRSB), superseding 
earlier versions by Ratra \& Peebles \shortcite{RPa} and by Kamionkowski et 
al. \shortcite{KRSS}.

The viability of the open CDM model has recently been investigated by Ratra 
\& Peebles \shortcite{RPa}. This paper predated the improved normalizations 
of inflationary open models to {\it COBE} supplied first by Kamionkowski et 
al. \shortcite{KRSS} and then more accurately again by GRSB (see also White 
\& Bunn 1995). Each of those collaborations provided only a brief 
account of the model's status against observations other than those of 
the microwave background. It is our aim in this paper to make a more 
extensive comparison of the model with observations. 

\section{The Open Universe Power Spectrum}

The model is defined by giving the spectrum ${\cal P}(k)$ of the density 
contrast $\delta$. Our approach towards constraining the model is to utilize 
linear perturbation theory, applied across as wide a range of scales as 
possible. By considering the formation of objects such as quasars and damped 
Lyman alpha systems at moderate redshift, it is possible to impose 
constraints on the spectrum at scales down to one megaparsec or less, while 
{\it COBE} probes scales of several thousands of megaparsecs, up to and even 
above the curvature scale. Between these extremes, a variety of different 
constraints can be applied.

All of the observations except the CMB anisotropy probe scales which are 
small compared with the Hubble distance, so we can use the Newtonian 
description of density perturbations to describe them. At any epoch well 
after matter domination sets in, the power spectrum of the density contrast 
is
\begin{equation}
\label{powerspec}
{\cal P}(k) = \delta_H^2 T^2(k)  \frac{g^2(\Omega)}{g^2(\Omega_0)} 
\left(\frac{k}{aH}\right)^4 \,.
\end{equation}
Here $a$ is the scale factor of the universe, $H=\dot a/a$ is the Hubble 
parameter (dots signifying time derivatives), $k$ is the comoving wavenumber 
(in inverse megaparsecs) and we are defining ${\cal P}(k)$ as the power per 
unit logarithmic interval of $k$. The transfer function $T(k)$ specifies the 
scale-dependent effect of the evolution of the density perturbation between 
horizon entry and matter domination, and is normalized to unity on large 
scales. The factor $g(\Omega)$ is introduced to allow for the growth law for 
perturbations in an open universe. It gives the total suppression of growth 
in an open universe relative to a critical density universe, and is 
accurately given\footnote{Numerical tests indicate that this fitting function 
is accurate to within one per cent for $\Omega_0$ of interest.} by the 
fitting function \cite{CPT}
\begin{equation}
\label{supp}
g(\Omega) = \frac{5}{2} \Omega \left[ 1 + \frac{\Omega}{2} + \Omega^{4/7} 
\right]^{-1} \,.
\end{equation}
Finally, the quantity $\delta_H$ specifies the overall normalization of the 
{\em present-day} spectrum. Its independence of $k$ indicates the assumption 
of an inflationary model leading to a scale-invariant spectrum; in typical 
inflationary models one would expect some deviation from this \cite{LL} and 
we shall consider this at the end of the paper. Inflationary models also 
predict the existence of gravitational waves, but their effect on the {\it 
COBE} normalization has not yet been successfully calculated and so we 
assume they are negligible. The value of $\delta_H$, when fixed by the {\it 
COBE} observations as discussed below, depends on $\Omega_0$, but it has only 
an extremely weak dependence on $H_0$ which can be comfortably ignored 
(throughout, a subscript 0 denotes the present day).

Many observations do not allow one to impose constraints on the power 
spectrum itself, but instead place constraints on the dispersion of the 
density contrast $\sigma(R)$ smoothed on a comoving scale $R$. We shall 
always use a top-hat filter $W(kR)$ defined by
\begin{equation}
\label{tophat}
W(kR) = 3 \left( \frac{\sin(kR)}{(kR)^3} - \frac{\cos(kR)}{(kR)^2} \right) 
\end{equation}
to perform the smoothing. The dispersion of the smoothed density contrast is 
easily calculated from a theoretical power spectrum as
\begin{equation}
\label{disp}
\sigma^2(R) = \int_0^\infty {\cal P}(k) \, W^2(kR) \, \frac{{\rm d}k}{k} \,.
\end{equation}
The prediction for the abundance of objects of various types has a very 
simple interpretation as a constraint on $\sigma(R)$, which cannot be easily 
represented as a power spectrum constraint.

Before proceeding to a full account of the data and their interpretation, let 
us be more specific regarding our assumptions. The parameters that we shall 
consider as freely variable are the total present density $\Omega_0$ and the 
present Hubble parameter $h$ (in units of $100 \; {\rm km} \, {\rm s}^{-1} \, 
{\rm Mpc}^{-1}$). An important contribution to be taken into account is the 
baryonic component of the density, $\Omega_{{\rm B}}$, which we take to be 
fixed by nucleosynthesis as $\Omega_{{\rm B}} h^2 = 0.0125$ 
\cite{NUCL}\footnote{We note that the more recent analysis of Copi, Schramm 
\& Turner \shortcite{CST} suggests that the traditional upper limit from 
Walker et al. \shortcite{NUCL} may be relaxed somewhat, though not 
sufficiently to impact on our results.}. In the presence of baryons, 
the usual scaling law of the transfer function with $\Omega_0 h$ (which is 
exact only for zero baryon density) can be replaced by an empirical scaling 
law with $\Omega_0 h \exp(- \Omega_{{\rm B}} -\Omega_{{\rm B}}/\Omega_0)$. 
This law was discovered by Sugiyama \shortcite{SUG94}, and generalizes a 
scaling law advertised by Peacock \& Dodds \shortcite{PD} to the case where 
$\Omega_0 < 1$. Although Sugiyama's calculations were made for the case of a 
flat universe with a cosmological constant, the difference between that and 
the present case only sets in long after the universe is matter dominated and 
so the shape is the same in our case. The different overall normalization 
of the spectrum between the two cases is of course included in the {\it COBE} 
normalization that we shall carry out.

We use the transfer function from Bardeen et al. \shortcite{BBKS},
\begin{eqnarray}
T_{{\rm CDM}}(q) & = & \frac{\ln \left(1+2.34q \right)}{2.34q} \times  
	\\ \nonumber
& & \hspace*{-1.7cm} 
	\left[1+3.89q+(16.1q)^2+(5.46q)^3+(6.71q)^4\right]^{-1/4} \,,
\end{eqnarray}
with $q = k/h\Gamma$, where the so-called `shape parameter' $\Gamma$ is 
defined as
\begin{equation}
\Gamma = \Omega_0 h \exp (-\Omega_{{\rm B}}-\Omega_{{\rm B}}/\Omega_0) \,,
\end{equation}
in accordance with Sugiyama \shortcite{SUG94} as discussed 
above\footnote{Note that Peacock \& Dodds \shortcite{PD} have a 
typographical error in the transfer function; we have confirmed that their 
results apply to the correct form.}.

Although $h$ is in principle freely variable, it is determined at some level 
of accuracy by the requirement of a reasonable fit to the galaxy correlation 
function (see further discussion below), which demands that $\Gamma$ should 
lie in the range $[0.22,0.29]$ at 95 per cent confidence level assuming a 
scale-invariant power spectrum \cite{PD}. Note though that, as $\Omega_0$ 
tends to 1, the required value of $h$ to achieve this begins to get 
uncomfortably small. 

Particularly for high $h$, one is in danger of a conflict between measured 
ages of stellar populations and the age of the universe. In an open universe 
the age is given by 
\begin{equation}
t_0 = \frac{1}{\Omega_0 H_0}\left[ \frac{\Omega_0}{1-\Omega_0} - 
\frac{\Omega_0^2}{2\left(1-\Omega_0\right)^{3/2}} \cosh^{-1} \left( 
\frac{2-\Omega_0}{\Omega_0} \right) \right] .
\end{equation}
If one fixes $H_0$, then higher ages are achieved by lowering $\Omega_0$. 
However, we have written it this way to emphasize an alternative view, which 
is that the galaxy correlation function more or less fixes (ignoring for 
now the baryonic corrections) the combination $\Omega_0 H_0$. Then the 
quantity in square brackets in the above formula is actually an {\em 
increasing} function of $\Omega_0$, peaking at $2/3$ when $\Omega_0 = 1$. 
Consequently, at fixed $\Gamma$, the desire for a large age favours a larger 
value of $\Omega_0$. To make this concrete, then fixing $\Gamma = 0.25$ 
taking the baryons into account gives the sample values $\Omega_0 = 0.2 
\Rightarrow h = 1.3 \Rightarrow \; {\rm Age} \; = 6$ Gyr; $\Omega_0 = 0.3 
\Rightarrow h = 0.89 \Rightarrow \; {\rm Age} \; = 9$ Gyr; $\Omega_0 = 0.4 
\Rightarrow h = 0.69 \Rightarrow \; {\rm Age} \; = 11$ Gyr; $\Omega_0 = 1.0 
\Rightarrow h = 0.32 \Rightarrow \; {\rm Age} \; = 20$ Gyr. In each case the 
15 per cent or so uncertainty in $\Gamma$ contributes a similar uncertainty 
to the age.

We shall adopt the extremely conservative view that the age should exceed 
$10$ Gyr, though there are many indications that the Universe is older 
(e.g. Demarque, Deliyannis \& Sarajedini 1991; Stockton, Kellogg \& Ridgway 
1995) which one can use to constrain cosmological parameters without 
reference to large scale structure.

\section{Normalization to {\it COBE}}

The most crucial piece of data is the overall normalization of the density
perturbation spectra, which we choose to match the microwave anisotropies at
large angular scales measured by the DMR experiment on the {\it COBE} 
satellite \cite{B94,Getal}. In the language of the usual spherical harmonic 
decomposition, {\it COBE} measures the multipoles with $l\la 30$, and for a 
given $\Omega_0$ the distance subtended at the surface of last scattering is 
comparable with the curvature if $l\la 2\sqrt{1-\Omega_0}/\Omega_0$. With the 
possible exception of the super-curvature modes defined below, this criterion 
gives an {\em upper bound} on the range of $l$ for which curvature can be 
significant. It may however be a considerable overestimate, because for 
$\Omega_0<1$ the dominant contribution to the CMB anisotropy can come from 
distances far closer than the surface of last scattering. In any case it 
allows curvature to affect only $l\la 6 $ even for $\Omega_0$ as low as 
$0.3$, which means that at most the lowest few multipoles of {\it COBE} are 
likely to be sensitive to curvature. 

To investigate the effect of curvature quantitatively, one must first ask how 
the Newtonian expression (\ref{powerspec}) should be continued to larger 
scales. As discussed in detail in Lyth \& Woszczyna \shortcite{LW}, a number 
of issues have to be addressed. 

First, in order to define the density contrast one has to specify a slicing 
of space-time into spatial hypersurfaces. We make the usual choice that the 
hypersurfaces are orthogonal to comoving observers, corresponding to what is 
called the `gauge invariant' density perturbation. In the era well after 
matter domination (which is the only one that concerns us) this is 
the same as the `synchronous gauge' density perturbation \cite{lystb}.

Secondly there is the definition of the spectrum. In discussing the 
stochastic properties of a given perturbation $f$, one assumes that it is a 
typical realization of a {\em random field} (an ensemble of functions 
together with a probability distribution for them). In both flat and curved 
space, the spectrum is defined with reference to an expansion in terms of 
eigenfunctions of the Laplacian, being the ensemble average of the modulus 
squared of the coefficient. Following Lyth \& Stewart \shortcite{lyst}, we 
denote the eigenvalue of the Laplacian by $-(k/a)^2$, and normalize the 
spectrum ${\cal P}_f(k)$ of a generic perturbation $f$ so that it gives the 
power per unit logarithmic interval of $k$. (By `power' we mean the ensemble 
mean square contribution to $f^2$, which is independent of position.) We are 
taking the random field to be Gaussian, which means that each coefficient has 
an independent Gaussian probability distribution, whose variance is 
essentially defined by the spectrum. (To make this statement precise one 
needs to take account of the fact that $k$ is a continuous, not a discrete, 
variable.)

Thirdly, there is the range of $k$ over which the spectrum is non-zero. If 
$k^{-1}$ is measured in units of the curvature scale 
$H_0^{-1}/\sqrt{1-\Omega_0}$, then it is known that the most general square 
integrable {\em function} can be constructed using only the eigenfunctions 
with $k^2>1$. For this reason, cosmologists have always assumed that the same 
is true for the most general Gaussian random field. That is, they have 
assumed that such a field can always be generated by keeping only the
the sub-curvature modes (those with $0<k^{-2}<1$) as distinct from
the super-curvature modes (those with $k^{-2}>1$). It has recently 
been pointed out \cite{LW} that this is not so; rather, mathematicians have 
known for half a century that in order to construct the most general Gaussian 
random field the spectrum (and therefore the eigenfunction expansion) needs 
to run over the full range $k^2>0$. Such super-curvature modes can arise in 
the single-bubble models of open inflation \cite{four}, though as we shall 
see it is a reasonable working hypothesis to assume that the effect of these 
is negligible.\footnote{A {\em smooth} continuation of the spectrum into the 
super-curvature regime would not have a significant effect on the CMB 
anisotropy \cite{LW}. One can also show that a delta function contribution at 
$k^2=0$ (the open universe Grishchuk-Zel'dovich effect) is not compatible 
with the data \cite{GLLW}.}

In most of the cosmology literature a different normalization of the 
spectrum is adopted, which is denoted by $P_f$ rather than by ${\cal 
P}_f$. In flat space, $P_f$ is normalized so that $k^3 P_f/(2\pi^2)$ is the 
power per unit logarithmic interval of $k$. Because super-curvature modes 
were never considered, this definition is customarily generalized to make 
$q^3P_f/(2\pi^2)$ the power per unit logarithmic interval of $q$, where 
$q^2=k^2-1$. (The motivation for considering $q^2$ instead of $k^2$ is 
that its range is $q^2>0$.) This leads to the relation 
\begin{equation}
P_f(q)=\frac{2\pi^2}{q(q^2+1)}{\cal P}_f(k(q)) \,.
\end{equation}

These preliminaries having been addressed, we are ready to ask what is the 
correct continuation to large scales of the flat space expression 
(\ref{powerspec})? A natural choice is to keep equation (\ref{powerspec}) as 
it stands, either retaining or dropping the super-curvature modes. If 
super-curvature modes are dropped there are other natural choices, based on 
the alternatively defined spectrum $P$ that we have just discussed. One can
take $P\propto q$ (the usual choice until recently) or $P\propto k$;
these choices multiply equation (\ref{powerspec}) by $(q/k)^2$ and $q/k$
respectively. Another choice \cite{KamS}, relating to the density 
contrast smoothed over a sphere of variable radius, makes $P\propto q^3$ for 
$q\la 1$ going smoothly over to $P\propto q$ for $q\ga 1$.

A further possibility is that, instead of focusing on the density contrast 
$\delta$, one can focus on the primordial curvature perturbation $\cal R$, 
given by \cite{Bardeen,lyst}
\begin{equation}
{\cal R}=\frac{5}{2} \left( \frac{a^2H^2}{3+k^2}\right) \delta \,.
\end{equation}
On small scales equation (\ref{powerspec}) corresponds to a 
scale-indep\-endent spectrum ${\cal P}_{\cal R}$. However, if ${\cal P}_{\cal 
R}$ is taken to be scale independent on large scales also, equation 
(\ref{powerspec}) is multiplied by a factor $[(3+k^2)/k^2]^2$. At $k^2=2$ 
this is a factor $\simeq 5$, corresponding to a factor $\sqrt 5$ in the rms 
perturbation, so it is more significant than the ambiguity associated 
with the definition of the spectrum, and the use of $q^2$ versus $k^2$. Thus 
the crucial decision is whether to regard the density perturbation or the 
curvature as the fundamental quantity.

The usual assumption that the perturbation originates as a vacuum 
fluctuation of the inflaton field decides in favour of the curvature,
because the inflaton field perturbation $\delta\phi$ is related to the 
curvature by ${\cal R}=(H/\dot \phi) \delta \phi$ which is scale-independent 
\cite{lyst,LL}. Making the arbitrary assumption of the conformal vacuum as 
the initial state in calculating the inflaton perturbation, the spectrum of 
$\delta \phi$ and therefore of $\cal R$ is flat \cite{lyst,RPb}. Recently 
it has been pointed out that in the bubble nucleation model the quantum 
fluctuation of the inflaton field, and hence the spectrum, is calculable 
without recourse to an arbitrary assumption concerning the initial vacuum 
state. According to Bucher et al. \shortcite{BGT} and Bucher \& Turok 
\shortcite{bt}, ${\cal P}_{\cal R}$ varies like $\coth(\pi q)$. At $k^2=2$ 
$(q^2=1)$ this factor is $1.06$, and even at $q^2=0.03$ it is only 2. In the 
single-bubble case, there is also the possibility of a discrete 
super-curvature mode \cite{five} provided the inflaton mass is light enough; 
however, again this should have only a small effect. Hence the difference 
between the conformal vacuum hypothesis and the bubble nucleation scenario is 
insignificant \cite{sasaki,bt}.\footnote{After this paper was accepted, 
Yamamoto \& Bunn \shortcite{YB} demonstrated explicitly that the 
normalization of the power spectrum in the single-bubble and conformal vacuum 
cases is nearly identical for any reasonable $\Omega_0$ (though the full 
details of the fit to {\it COBE} differ somewhat between the two cases).}

Although present versions of the open inflationary scenario are not without 
their problems, the situation does appear to have improved recently, in that 
the fine-tuning of initial conditions required in early models \cite{lyst} is 
not necessary in the bubble nucleation model. At present it does however 
still seem necessary to have some fine-tuning in the parameters of these 
models \cite{LinMez}. Despite this, open inflation models are the natural 
underpinning of structure formation models in an open universe.

The upshot of the above discussion is that the criterion of some smooth 
continuation suggests a moderate amount of ambiguity in the large-scale power 
spectrum, which is however practically eliminated if the perturbation 
originates as a quantum fluctuation of the inflaton field. According to 
Sugiyama \& Silk \shortcite{SSilk}, even the moderate ambiguity suggested by 
smoothness is not very significant, because it affects only the low CMB 
multipoles which are poorly determined because of cosmic variance.
However, in this paper we adopt for definiteness the hypothesis of a flat 
curvature spectrum, which is the prediction of inflation.

Given the spectrum, one can calculate the expected values for the multipoles 
measured by {\it COBE}, and compare with observations to determine the 
best-fitting normalization. The calculation is substantially more complex 
than that for critical density models, which is almost analytic, and in the 
literature it has been developed in several stages. In this paper we use the 
most recent and sophisticated determination, given by GRSB. They take the 
curvature perturbation spectrum to be flat, and fit the full spectrum of 
anisotropies including the Doppler peak using a method based on Fourier 
analysis on the cut sky for which {\it COBE} data are available. This 
normalization is more sophisticated than that of Kamionkowski et al. 
\shortcite{KRSS}, who normalized to the $10 \degr$ variance \cite{B94} with 
a correction incorporated for the beam profile and non-orthogonality of the 
monopole and dipole subtraction \cite{W94}. They also arbitrarily increased 
the error bar to 30 per cent. 

The outcome of the GRSB analysis is that essentially all values of $\Omega_0$ 
are capable of providing an acceptable fit to the {\it COBE} data for a 
suitable choice of normalization. They do not explicitly state the 
normalization of the power spectrum that they get for each $\Omega_0$. 
However, they do give values of $\sigma(8 \, h^{-1} {\rm Mpc})$ for specific 
choices of $h$, directly calculated from their Boltzmann code. We use these 
to calculate the large-scale normalization of the power spectrum 
($\delta_H(\Omega_0)$ in equation (\ref{powerspec})), which is {\em 
independent} of $h$. This can then be used to calculate $\sigma(R)$ using 
equation (\ref{disp}) for any value of $h$ by using the appropriate transfer 
function.

We find that the normalization can be accurately represented, to within 2 
per cent for $0.1 < \Omega_0 < 1$, by the fitting function
\begin{equation}
\label{COBEnorm}
\delta_H^2(\Omega_0) = \left( 4.10 + 8.83 \Omega_0 - 8.50 \Omega_0^2 
\right)\times 10^{-10} \,,
\end{equation}
where we use the GRSB analysis which includes the quad\-ru\-pole (almost no 
change arises if the quadrupole is dropped from the analysis).

The normalization from GRSB has an error bar of 8 per cent (more or less 
independently of $\Omega_0$), as compared with the 30 per cent used by 
Kamionkowski et al. \shortcite{KRSS}. Although this tighter error bar is 
certainly more constraining, this normalization is quite a bit higher than 
used by Kamionkowski et al. An increase in the normalization generally acts 
in favour of the lower density models when it comes to comparing with the 
observations. As the {\it COBE} normalization has a considerably smaller 
error bar than other observations, on occasion we shall take this 
normalization as fixed, ignoring its error bar.

\section{Smaller scale constraints}

A wide range of observations provide a variety of constraints on the power 
spectrum on scales of order 1 to 100\,Mpc. These include the distribution of 
galaxies and clusters, the peculiar motions of galaxies and the abundances of 
various objects including clusters, quasars and damped Lyman alpha systems. 
Our approach is to use only the most powerful ones, as described elsewhere 
for the case $\Omega_0 = 1$ (Liddle \& Lyth 1995; Liddle et al., in 
preparation). When the spatial geometry is changed, all constraints need to 
be recalculated for a variety of reasons, amongst which the primary ones are 
a suppressed rate of perturbation growth at low redshift and an amended 
relation between scale and mass. 

\subsection{The galaxy correlation function}

One of the most highly advertised problems with the standard cold dark matter 
scenario is its failure to reproduce correctly the shape of the galaxy 
correlation function on scales of tens of megaparsecs, on the reasonable 
assumption of a scale-independent bias parameter for galaxies of a given 
type. For CDM models, this is quantified via the shape parameter $\Gamma$ 
which we have already introduced, and from a detailed analysis of a 
compilation of data sets Peacock \& Dodds \shortcite{PD} obtain the very 
stringent constraint $\Gamma = 0.255^{+0.038}_{-0.033}$ at the 95 per cent 
confidence limit assuming a scale-invariant power spectrum. Provided one is 
willing to tolerate a sufficiently small $h$ (around $0.35$), the shape 
parameter can be fitted in a critical density universe. 

In addition to providing a constraint on the shape parameter, the galaxy 
distribution data also in principle constrain the normalization of the 
spectrum through redshift space distortions and non-linear effects. By 
choosing a scale in the middle of the data the best-fit amplitude can be 
found independently of $\Gamma$; assuming $\Omega_0=1$ and $b_{I}=1$, where 
$b_{I}$ is the bias parameter for {\it IRAS}-selected galaxies, we find the 
constraint $\sigma(15.1 \, h^{-1} {\rm Mpc})=0.40 \pm 0.03$. For 
general $\Omega_0$, Peacock \& Dodds provide a best-fitting bias parameter, 
and by fitting for this and processing through the redshift distortion factor 
for general $\Omega_0$ one obtains the formal result with 1$\sigma$ error 
(almost entirely due to the uncertainty in the bias) of
\begin{equation}
\sigma(15.1 h^{-1} {\rm Mpc}) = (0.40 \pm 24 {\rm per \; cent}) 
f(\Omega_0)\,,
\end{equation}
where the fitting function $f(\Omega_{0})$ is given by
\begin{equation}
f(\Omega_{0})=1.62+0.81\Omega_{0}-2.60\Omega_{0}^{2}+1.31\Omega_{0}^{3}\,.
\end{equation}
However, the literature contains a widespread range of estimates of the bias 
parameter (see for example the compilation in Dekel \shortcite{Dekel}), 
suggesting a true uncertainty larger than that advertised by Peacock \& 
Dodds. As this result is anyway less constraining than other data, we choose 
not to impose this constraint.

A chi-squared analysis of the sixteen data points in table 1 of Peacock \& 
Dodds \shortcite{PD}, taking $\Gamma$ and the normalization as fitting 
parameters, has 14 degrees of freedom. Unfortunately the minimum chi-squared 
is somewhat low (about 12). It is perfectly reasonable that this has occurred 
by chance, though it could also have an origin in weak correlations of 
neighbouring data or through non-normal errors. This prevents us 
incorporating their full data set into a chi-squared analysis along with 
other data, because such an analysis allows other data points to receive a 
high chi-squared in compensation because their data set has so many more 
points. We have tried to evade this by only incorporating the shape parameter 
into a chi-squared test on all the data. 

\subsection{Bulk flows and POTENT}

For a given present-day amplitude of density perturbations, the predicted 
peculiar velocities depend quite strongly on the value of $\Omega_0$, 
becoming much smaller in the low-density case. The best measurements of the 
bulk flow available are those found via the POTENT technique of velocity 
field reconstruction \cite{BD89}. For the Mark III data set \cite{Dekel}, the 
velocity has been evaluated in spheres about our position for a range of 
radii. However, these separate determinations are not independent as the rms 
bulk flow is sensitive to long wavelengths to a much greater extent than the 
density contrast. We therefore concentrate on a single measurement, which is 
the bulk flow smoothed on a scale of $40 \, h^{-1}$ Mpc. The method used to 
generate this requires an additional Gaussian smoothing on $12 \, h^{-1}$ Mpc 
in order to generate the original continuous velocity field used as a 
starting point. The theoretical prediction for the rms bulk flow is therefore 
given by 
\begin{equation}
\sigma_v^2(R) = H_0^2 \Omega_0^{1.2} \! \int_0^\infty \! \! W^2(kR) \exp 
	\left(-(12\, h^{-1} k)^2 \right)\, \frac{{\cal P}_0}{k^2} 
	\frac{{{\rm d}}k}{k} ,
\end{equation}
where $W(kR)$ is the top-hat window given by equation (\ref{tophat})
and the factor $\Omega_0^{1.2}$ is an extremely accurate fitting function to 
the $\Omega_0$-dependent velocity suppression.

The Mark III POTENT analysis gives the bulk flow in a $40 \, h^{-1}$ Mpc 
sphere as \cite{Dekel}
\begin{equation}
\label{bf}
v_{{\rm obs}}(40 \, h^{-1} {\rm Mpc}) = 373 \pm 50 \; {\rm km} \, 
	{\rm s}^{-1}\,,
\end{equation}
and this provides the best estimate of $\sigma_v(40 \, h^{-1} {\rm Mpc})$. 
The error given in expression (\ref{bf}) arises from different ways of 
dealing with sampling-gradient bias and can thus be thought of as reflecting 
the systematic uncertainty in the POTENT analysis. Additionally there is an 
intrinsic uncertainty in the POTENT calculation due to random distance 
errors, which at the 1$\sigma$ level is $\simeq 15$ per cent \cite{Dekel}. 
The observational error is dominated by cosmic variance; since the mean 
square bulk flow is the sum of the squares of the three velocity components, 
each of which is Gaussian distributed, it follows a $\chi^2$ distribution 
with three degrees of freedom. This enables a calculation of the cosmic 
variance error in using the bulk flow as an estimator of the normalization of 
the dispersion of the density contrast, that error being $89$ per cent 
upwards and $24$ per cent downwards at the $68$ per cent confidence level 
which notionally corresponds to 1$\sigma$. At the $95$ per cent confidence 
level the error bars are $+273$ per cent and $-43$ per cent. We can improve 
on this by modelling the observational errors and convolving with the 
theoretical distribution. Assuming that the error in expression (\ref{bf}) 
corresponds to something like $95$ per cent confidence (though as it is the 
smallest error this assumption is insignificant), then the convolution of the 
three types of error results in the total error in using the Mark III POTENT 
bulk flow calculation as an estimator of the normalization of the dispersion 
of the density contrast. The increase in the error range as compared with 
cosmic variance alone is not large, the total error range being $+98$ per 
cent to $-25$ per cent at the $68$ per cent confidence level, and $+295$ per 
cent to $-47$ per cent at the 95 per cent confidence level. Only the lower 
limits are useful for us.

We note that a constraint on the value of $\Omega_0$ can be extracted 
from non-linear effects on the peculiar velocities, yielding $\Omega_0 \ga 
0.3$ at least at the 2$\sigma$ confidence level \cite{Dekel} which serves to 
reinforce our conclusions.

The scale at which the bulk flow data apply is of order 1 per cent of the 
Hubble distance, so one might wonder if general relativistic effects might be 
detectable. The formulae that we have given remain valid in that case, 
provided that the density perturbation is defined on hypersurfaces orthogonal 
to comoving observers, and that the peculiar velocity is defined with respect 
to worldlines having zero shear \cite{brly}. It is noted in GRSB that, with a 
different choice, the theoretically calculated bulk flow is different by 
several per cent, which is not totally insignificant. This suggests that a 
careful analysis of the observations using general relativity would be 
worthwhile, using a specific set of worldlines to define the peculiar 
velocity. In any event, as long as Newtonian physics is used to analyse the 
data there is certainly no point in going beyond that framework in the 
theoretical calculation.

\subsection{Object abundances}

In the case of a critical density universe the standard analytical technique 
to calculate object abundances relies on the use of the Press--Schechter 
theory \cite{PS}, which has been found through $N$-body simulations to 
provide a good approximation \cite{LC}. This kind of comparison between 
analytical techniques and $N$-body simulations has not been performed to the 
same extent for an open universe. However, the derivation of the 
Press--Schechter theory relies solely on statistical arguments; there is 
nothing in it that explicitly relies on the background cosmology. It should 
therefore also be applicable in an open universe. We shall use it to obtain 
constraints on the abundances of galaxy clusters and damped Lyman alpha 
systems.

Using the Press--Schechter theory, the fraction of the matter in the universe 
that is in collapsed objects above a given mass at a redshift $z$ is given 
simply by 
\begin{equation}
\label{PS}
\frac{\Omega(>M(R),z)}{\Omega(z)} = {\rm erfc} \left( \frac{\delta_{{\rm 
c}}}{\sqrt{2} \, \sigma(R,z)} \right) \,,
\end{equation}
where $\delta_{{\rm c}}$ is the threshold value fixed by comparison with 
$N$-body simulations, $\sigma(R,z)$ is the dispersion smoothed on scale $R$ 
at redshift $z$ and `erfc' is the complementary error function. The 
appropriate value for $\delta_{{\rm c}}$ in this expression depends on the 
type of collapse one wants to consider, and on the type of filter one uses to 
carry out the smoothing. In a critical density universe the spherical 
collapse of a top-hat perturbation is associated with $\delta_{{\rm c}}=1.7$. 
Non-spherical collapse along all three axes of symmetry is associated with 
higher values for $\delta_{{\rm c}}$, whilst non-spherical collapse along the 
first and second collapsing axes is associated with smaller values (e.g. 
Monaco 1995). As the value of $\delta_{{\rm c}}$ is determined by the 
time-scale of collapse of a given type of perturbation, one might expect it 
to be quite sensitive to the background cosmology being considered. However, 
this does not seem to be the case when one moves from a critical density 
universe to an open universe. Lilje \shortcite{lilje}, Lacey \& Cole 
\shortcite{LC93} and Colafrancesco \& Vittorio \shortcite{CV} found that, at 
least for any type of collapse where $\delta_{{\rm c}}\leq1.7$ in a critical 
density universe, the value of $\delta_{{\rm c}}$ varies at most by $5$ per 
cent when one goes from an $\Omega_0 = 1$ universe to one with 
$\Omega_{0}=0.1$. This applies at the present epoch, therefore implying that 
the same change in background cosmology will give rise to an even smaller 
variation in $\delta_{{\rm c}}$ at higher redshifts, as presently open 
universes approach flatness with increasing redshift. 

The abundance of galaxy clusters is used to constrain the present-day power 
spectrum. In order to constrain shorter scales, which are well into the 
non-linear regime today, a successful technique is to study objects at high 
redshift, when those scales were still in the linear regime. By using linear 
theory to scale those constraints to the present, one can compare directly 
with the present-day predicted linear power spectrum. The most useful objects 
on which data are available are the damped Lyman alpha systems (Lanzetta, 
Wolfe \& Turnshek 1995; Storrie-Lombardi et al. 1995). These offer a tighter 
constraint than the quasar abundance, the latter being weakened by unknown 
efficiency factors such as the required number of generations of quasars, and 
by the uncertainty in the required host galaxy mass.

We wish to take into account the growth of density perturbations between a 
redshift, say, around four and the present. As $\Omega_0$ is decreased, the 
amount of growth between these epochs becomes highly suppressed, which is one 
of the main reasons why the present normalization of the primordial spectrum 
is lower than in the critical density case. On the other hand, this effect 
helps with high-redshift object formation since, for a given present-day 
normalization, the perturbations at high redshift are substantially higher 
than if the universe were flat.

In a critical density matter-dominated universe, $\sigma(M,z)$ simply grows 
proportionally to $(1+z)^{-1}$. In an open universe, there is a suppression 
$g$ in growth relative to this given by equation (\ref{supp}). This equation 
can be applied at any epoch, using the redshift dependence of $\Omega$ which 
in a matter-dominated universe is given by
\begin{equation}
\Omega(z) = \Omega_0 \, \frac{1+z}{1+\Omega_0 z} \,.
\end{equation}
One therefore needs to apply the growth factor for a critical density 
universe, correcting for the suppression both at the redshift of the 
observation and at the present, to get a constraint on the present-day power 
spectrum from
\begin{equation}
\label{growth}
\sigma(M,z=0) = \sigma(M,z) \, (1+z) \frac{g(\Omega_0)}{g(\Omega(z))} \,.
\end{equation}
  
\subsubsection{Cluster abundance}
 
A large galaxy cluster has a typical mass of about $10^{15} \, {{\rm 
M}}_{\sun}$, which corresponds to a linear scale of around $8 \, h^{-1}$ Mpc. 
Such clusters are relatively rare, indicating that this scale is still in the 
quasi-linear regime. One is then able to use the Press--Schechter theory to 
calculate $\sigma_{8}\equiv\sigma(8 \, h^{-1} {\rm Mpc})$. To our knowledge 
the first to attempt this was Evrard \shortcite{E89}, followed by Henry \& 
Arnaud \shortcite{HA}. Both these analyses were only valid for a critical 
density CDM universe, and though using different observations they reached 
essentially the same result. Then White, Efstathiou \& Frenk \shortcite{WEF} 
again obtained a result similar to the previous two, and extended the 
analysis to a flat CDM universe with non-zero cosmological constant. Our 
analysis is similar to theirs extended to an open universe, the main 
difference being that we attempt to take into account that clusters with 
equal mass which virialize at different redshifts have distinct properties, 
like velocity dispersion and X-ray temperature, at the present.   

A variety of different observations are available concerning the abundance of 
clusters. To use the Press--Schechter theory, it is vital to have good mass 
estimates as well as an estimate of the number density. Galaxy cluster 
catalogues assembled through optical selection from photographic plates, 
even disregarding the subjective nature of such selection, suffer from 
possible errors in cluster identification due to foreground and background 
contamination in the galaxy counts. Furthermore, the velocity dispersion, the 
optical observable most directly related to the cluster mass, is prone to 
serious projection effects and possible velocity biases. On the other hand, 
cluster identification through X-ray emission is free from foreground and 
background contamination, as X-rays are only produced in deep potential 
wells, and the X-ray observable most directly associated with the cluster 
mass, the mean X-ray temperature, is only very weakly affected by 
projection effects. Accordingly, we choose to use X-ray instead of optical 
data. 

The observed number density of clusters per unit temperature, $n(k_{{\rm 
B}}T)$, at $z=0$ was calculated by Henry \& Arnaud \shortcite{HA}. They found 
that clusters with a mean X-ray temperature of 7 keV have a present number 
density 
\begin{equation}
\label{nd}
n(7 \; {\rm keV} , 0) = 2.0_{-1.0}^{+2.0}\times10^{-7}h^{3} 
\; {\rm Mpc}^{-3} \; {\rm keV}^{-1} \,.
\end{equation}

The comoving number density of clusters in a mass interval ${{\rm d}}M_{{\rm 
v}}$ about virial mass $M_{{\rm v}}$ at a redshift $z$ is obtained by 
differentiating equation (\ref{PS}) with respect to the mass and multiplying 
it by $\rho_{{\rm b}}/M_{{\rm v}}$, where $\rho_{{\rm b}}$ is the comoving 
background density (a constant during matter domination), thus giving
\begin{eqnarray}
\label{mfa}
n(M_{{\rm v}},z) \, {{\rm d}} M_{{\rm v}} = & & \\ \nonumber 
& & \hspace*{-2cm} -\sqrt{\frac{2}{\pi}} \frac{\rho_{{\rm b}}}{M_{{\rm v}}}
	\frac{\delta_{{\rm c}}}{\Delta^{2}(z)}
	\frac{{\rm d}\Delta(z)}{{\rm d}M_{{\rm v}}}\exp \left[
	-\frac{\delta_{{\rm c}}^{2}}{2\Delta^{2}(z)}\right] \, 
	{{\rm d}}M_{{\rm v}} \,,
\end{eqnarray}
where $\Delta\equiv\sigma(r_{{\rm L}})$ with $r_{{\rm L}}$ the comoving 
linear scale associated with $M_{{\rm v}}$, $r_{{\rm L}}^3=3M_{{\rm 
v}}/4\pi\rho_{{\rm b}}$. Traditionally the cluster abundance is used to 
constrain the dispersion at $8 \, h^{-1}$ Mpc, and the quantity $\Delta$ is 
specified by an analytic approximation to the power spectrum in the vicinity 
of this scale. Generally, one can write
\begin{equation}
\label{del}
\Delta(z)=\sigma_{8}(z) \left(\frac{r_{{\rm L}}}{8\,h^{-1} \; 
	{\rm Mpc}}\right)^{-\gamma(r_{{\rm L}})} \,.
\end{equation}
For the CDM spectra we adopt the form (for $n = 1$)
\begin{equation}
\gamma(r_{{\rm L}})=(0.3 \Gamma + 0.2)\left[2.92 + \log\left(
	\frac{r_{{\rm L}}}{8\, h^{-1}\; {\rm Mpc}}\right)\right]\,.
\end{equation}
This is a more sophisticated analytic approximation than the power-law 
approximation used by White et al. \shortcite{WEF}; the open universe 
calculation requires accuracy over a wider range of scales (note also that 
their $\Gamma$ has a slightly different definition). This approximation is 
accurate to within $1$ per cent for $r_{{\rm L}}$ within a factor of 4 of 
$8\, h^{-1}\; {\rm Mpc}$ for the $\Gamma$ values in which we are primarily 
interested.

Note that, in any CDM model, $\gamma(r_{{\rm L}})$ is redshift independent 
since the growth of perturbations is independent of scale. Using expression 
(\ref{del}) to calculate the derivative in equation (\ref{mfa}) we therefore 
get
\begin{eqnarray}
\label{mf}
n(M_{{\rm v}},z) \, {{\rm d}}M_{{\rm v}} = & & \\ \nonumber
& & \hspace*{-2cm} \sqrt{\frac{2}{\pi}}\frac{\rho_{{\rm b}}}{M_{{\rm v}}^{2}}
	\frac{2.92(0.3\Gamma+0.2)\delta_{{\rm c}}}{3\Delta(z)}
	\exp\left[-\frac{\delta_{{\rm c}}^{2}}{2\Delta^{2}(z)}\right] 
	\, {{\rm d}}M_{{\rm v}} \,.
\end{eqnarray}

As large clusters are relatively rare, it is reasonable to assume that shear 
did not play an important part during their collapse, which to a good 
approximation can then be considered to have occurred spherically \cite{Ber}. 
Nevertheless, we shall include an assumed 1$\sigma$ dispersion of $\pm 0.1$ 
in the value of $\delta_{{\rm c}}$. Bearing in mind that varying the 
background cosmology has a negligible effect on the value of $\delta_{{\rm 
c}}$ we then use $\delta_{{\rm c}} = 1.7 \pm 0.1$ when needed for all our 
models at all $z$.

For the type of models we are considering, Hanami \shortcite{H} has shown 
that 
\begin{equation}
\label{mvprop}
M_{{\rm v}} \propto \Omega_{0}^{-1/2} \, \chi^{-1/2}(1+z_{{\rm c}})^{-3/2}
	(k_{{\rm B}}T)^{3/2}h^{-1} \,,
\end{equation}
where  
\begin{equation}
\chi = 1 + (\Omega_{0}^{-0.8}-1)
	(1+\Omega_{0}^{0.5}z_{{\rm m}})^{-\Omega_{0}^{-0.4}}\,.
\end{equation}
Here $z_{{\rm c}}$ and $z_{{\rm m}}$ are the redshifts of cluster 
virialization and turnaround respectively; they are related by the expression 
$(1+z_{{\rm m}}) \simeq 2^{2/3} (1+z_{{\rm c}})$. The scalings in equation 
(\ref{mvprop}) have been found through hydrodynamical $N$-body simulations to 
hold remarkably well in a $\Omega_{0}=1.0$ CDM model (Navarro, Frenk \& White 
1995).

In order to normalize equation (\ref{mvprop}) we use results from the 
hydrodynamical $N$-body simulations for a $\Omega_{0}=1.0$ CDM model 
performed by White et al. \shortcite{WNEF}. From a catalogue of 12 simulated 
clusters with a wide range of X-ray temperatures they estimated that a 
cluster with a present mean X-ray temperature of 7.5 keV corresponds to a 
mass within one Abell radius (1.5 $h^{-1}$ Mpc) of the cluster centre of 
$M_{{\rm A}}=(1.10 \pm 0.22)\times10^{15} \, h^{-1} \; {{\rm M}}_{\sun}$. The 
error arises from the dispersion in the catalogue and is supposed to 
represent the 1$\sigma$ significance level. White et al. \shortcite{WNEF} 
also found that the simulated clusters had a density profile in their outer 
regions approximately described by ${\rho}_{{\rm c}}(r) \, \propto \, r^{-2.4 
\pm 0.1}$. This same result was obtained by Metzler \& Evrard \shortcite{ME} 
and Navarro et al. \shortcite{NFW}. Bearing in mind that the cluster virial 
radius in a $\Omega_{0}=1.0$ universe encloses a density 178 times the 
background density, it is then straightforward to calculate the cluster 
virial mass from $M_{{\rm A}}$. Through a Monte Carlo procedure, where we 
assume the errors in $M_{{\rm A}}$ and in the exponent of ${\rho}_{{\rm 
c}}(r)$ to be normally distributed, we find $M_{{\rm v}}=(1.23 \pm 
0.32)\times10^{15} \, h^{-1} \; {{\rm M}}_{\sun}$ for a cluster with a 
present mean X-ray temperature of 7.5 keV in a $\Omega_{0}=1.0$ universe. 
Assuming that such a cluster virialized at a redshift of $z_{{\rm c}} \simeq 
0.05 \pm 0.05$ (e.g Metzler \& Evrard 1994; Navarro et al. 1995), 
we can now normalize equation (\ref{mvprop}):
\begin{eqnarray}
M_{{\rm v}} & = &(1.32 \pm 0.34) \times 10^{15} \times \\ \nonumber 
& & \hspace*{-0.5cm} \Omega_{0}^{-1/2} \, 
\chi^{-1/2}(1+z_{{\rm
	c}})^{-3/2}\left(
	\frac{k_{{\rm B}}T}{7.5\,{\rm keV}}\right)^{3/2} 
	\, h^{-1} \; {{\rm M}}_{\sun} \,.
\end{eqnarray}
This result is in very close agreement with the one obtained by Evrard 
\shortcite{E} from his own hydrodynamical $N$-body simulations. Hence the 
virial mass $M_{{\rm v}}$ for a cluster with a present mean X-ray 
temperature of 7 keV is given by  
\begin{equation}
\label{mv}
M_{{\rm v}} = (1.2 \pm 0.3) \times10^{15} \, \Omega_{0}^{-1/2} 
	\, \chi^{-1/2}(1+z_{{\rm c}})^{-3/2} \, h^{-1} \; {{\rm M}}_{\sun}\,.
\end{equation}

As one can see from equation (\ref{mvprop}), the relation between the cluster 
virial mass and its mean X-ray temperature depends on the redshift of 
cluster virialization. One therefore expects that at the present there will 
be some dispersion in the virial masses of clusters with the same mean 
X-ray temperature. This dispersion increases with decreasing $\Omega_{0}$, 
as, due to the slower growth of density perturbations in lower $\Omega_{0}$ 
models, cluster formation at a given scale proceeds over a greater redshift 
interval. 

According to Press--Schechter theory the comoving number density of clusters 
with virial mass in an interval ${{\rm d}} M_{{\rm v}}$ at $M_{{\rm v}}$ 
which virialize in a redshift interval ${{\rm d}}z$ at redshift $z$ and 
survive until the present is given by \cite{Sa}
\begin{eqnarray}
\label{ncom}
N(M_{{\rm v}},z) \, {{\rm d}}M_{{\rm v}} \, {{\rm d}}z = & & \\ \nonumber
& & \hspace*{-2cm} \left[-\frac{\delta_{{\rm c}}^2}{\Delta^2(z)}
	\frac{n(M_{{\rm v}},z)}{\sigma_8(z)}
	\frac{{\rm d}\sigma_8(z)}{{\rm d}z}\right]
	\frac{\sigma_8(z)}{\sigma_8(z=0)} \,  {{\rm d}}M_{{\rm v}} \, 
	{{\rm d}}z \,,
\end{eqnarray}
where $\sigma_8(z)$ and ${{\rm d}}\sigma_8(z)/{{\rm d}}z$ are calculated 
using equation (\ref{growth}). In equation (\ref{ncom}) the expression within 
the square brackets gives the formation rate of clusters with virial mass 
$M_{{\rm v}}$ at redshift $z$, whereas the fraction outside gives the 
probability of these clusters surviving until the present. If one now assumes 
that at each redshift $z$ the cluster virial mass $M_{{\rm v}}$ in equation 
(\ref{ncom}) is determined by expression (\ref{mv}) with $z_{{\rm c}}=z$, 
then equation (\ref{ncom}) gives the comoving number density of clusters per 
unit mass which virialize at each redshift $z$ and survive up to the present 
such that they have a mean X-ray temperature of 7 keV at the present. Through 
the chain rule we can then determine the comoving number density of clusters 
per unit temperature that virialize at each redshift $z$ and survive up to 
the present such that they have a mean X-ray temperature of 7 keV at the 
present
\begin{eqnarray}
N(k_{{\rm B}}T,z) \, {{\rm d}} (k_{{\rm B}}T) \, {{\rm d}}z & =  &  
	\frac{{\rm d}M_{{\rm v}}}{{\rm d}(k_{{\rm B}}T)}
	N(M_{{\rm v}},z) \, {{\rm d}}(k_{{\rm B}}T) \, {{\rm d}}z \,,
	\nonumber \\ 
 & = & \frac{3}{2} \frac{M_{{\rm v}}}{k_{{\rm
	B}}T}N(M_{{\rm v}},z) \, {{\rm d}}(k_{{\rm B}}T) \, {{\rm d}}z \,,
\end{eqnarray}
where the second equality uses equation (\ref{mvprop}). We therefore have 
\begin{eqnarray}
\label{nt}
N(k_{{\rm B}}T,z) \, {{\rm d}}(k_{{\rm B}}T) \, {{\rm d}}z = & & \\ \nonumber
& & \hspace*{-2.5 cm} -\frac{3}{2}\frac{M_{{\rm v}}}{k_{{\rm B}}T} 
	\frac{\delta_{{\rm c}}^{2}}{\Delta^{2}(z)}
	\frac{n(M_{{\rm v}},z)}{\sigma_{8}(z=0)}
	\frac{{\rm d}\sigma_{8}(z)}{{\rm d}z} \, {{\rm d}}(k_{{\rm B}}T) 
	\, {{\rm d}}z \,.
\end{eqnarray}

Numerically integrating this expression from $z=0$ to $z=\infty$ then gives 
the present comoving number density of clusters per unit temperature with a 
mean X-ray temperature of 7 keV as a function of $\Omega_{0}$ and of the 
present value of $\sigma_{8}$. Comparing with the observational value given 
by equation (\ref{nd}) we thus get $\sigma_{8}(\Omega_{0})$. We find to a 
good approximation that 
\begin{equation}
\label{final}
\sigma_{8} = \left( 0.60 ^{+32 \; {{\rm per \; cent}}}_{-24 \; {{\rm per 
\;
	cent}}}	\right)	{\Omega_{0}^{-C(\Omega_{0})}}\,,
\end{equation}
where $C(\Omega_{0})=0.37+0.13\Omega_{0}-0.02\Omega_{0}^{2}$ is a fitting 
function representing the changing power-law index of the $\Omega_0$ 
dependence. We have computed the uncertainty using a Monte Carlo method; it
arises from the dispersions in the observational value of $\Gamma$, in the 
assumed value for $\delta_{{\rm c}}$, and in expressions (\ref{nd}) and 
(\ref{mv}). The confidence level quoted in equation (\ref{final}) is 95 per 
cent. Further details of the calculation of the uncertainty will be provided 
elsewhere \cite{VL}. 

\subsubsection{Damped Lyman alpha systems}

Many types of model with $\Omega_0=1$, such as mixed dark matter models, are 
strongly constrained by data on damped Lyman alpha systems 
\cite{MM,KC,MB,KDLAS}. However, the constraint becomes weaker as $\Omega_0$ 
is reduced, as we will now see.

Instead of the widely quoted data of Lanzetta et al. \shortcite{LWT}, we use 
the recent data of Storrie-Lombardi et al. \shortcite{storrie} which revises 
downwards\footnote{Note that this still ignores the effect of gravitational 
lensing, which it is claimed can reduce the estimated abundance by a further 
50 per cent \cite{bartelmann}.} the estimated abundances at a redshift of 
around 3 and provides a new estimate at redshift 4. The strongest constraint 
comes from the redshift 4 point, and so we shall concentrate on it. However, 
the constraint is not significantly weakened if the redshift 3 point is used 
instead, and in any case we shall see that these data are not very 
constraining for open CDM models.

\begin{figure*}
\centering
\leavevmode\epsfysize=10.0cm \epsfbox{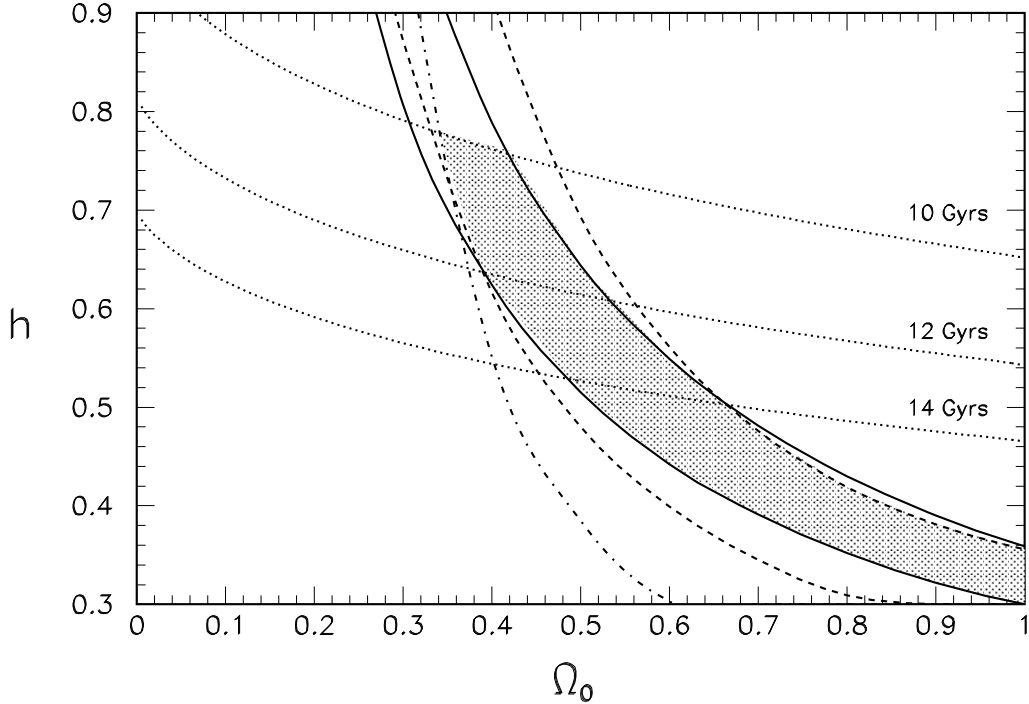}
\caption[figure1]{The constraints plotted in the 
$\Omega_0$--$h$ plane, assuming baryon density as given by standard 
nucleosynthesis. All models are normalized to the {\it COBE} data as given by 
GRSB, and the constraints shown are all at 95 per cent confidence. The solid 
lines are limits on the shape parameter. The dashed lines are limits from 
the cluster abundance. The dot-dashed line is the {\em lower} limit from 
POTENT (since it comes from confidence intervals, formally it is $97.5$ per 
cent as a lower limit, the upper limit not being shown here). Finally, the 
labelled dotted lines are contours of constant age as indicated. The shaded 
region shows the parameter space not excluded at greater than 95 per cent on 
any single piece of data.}
\end{figure*}

\begin{figure*}
\centering
\leavevmode\epsfysize=10.0cm \epsfbox{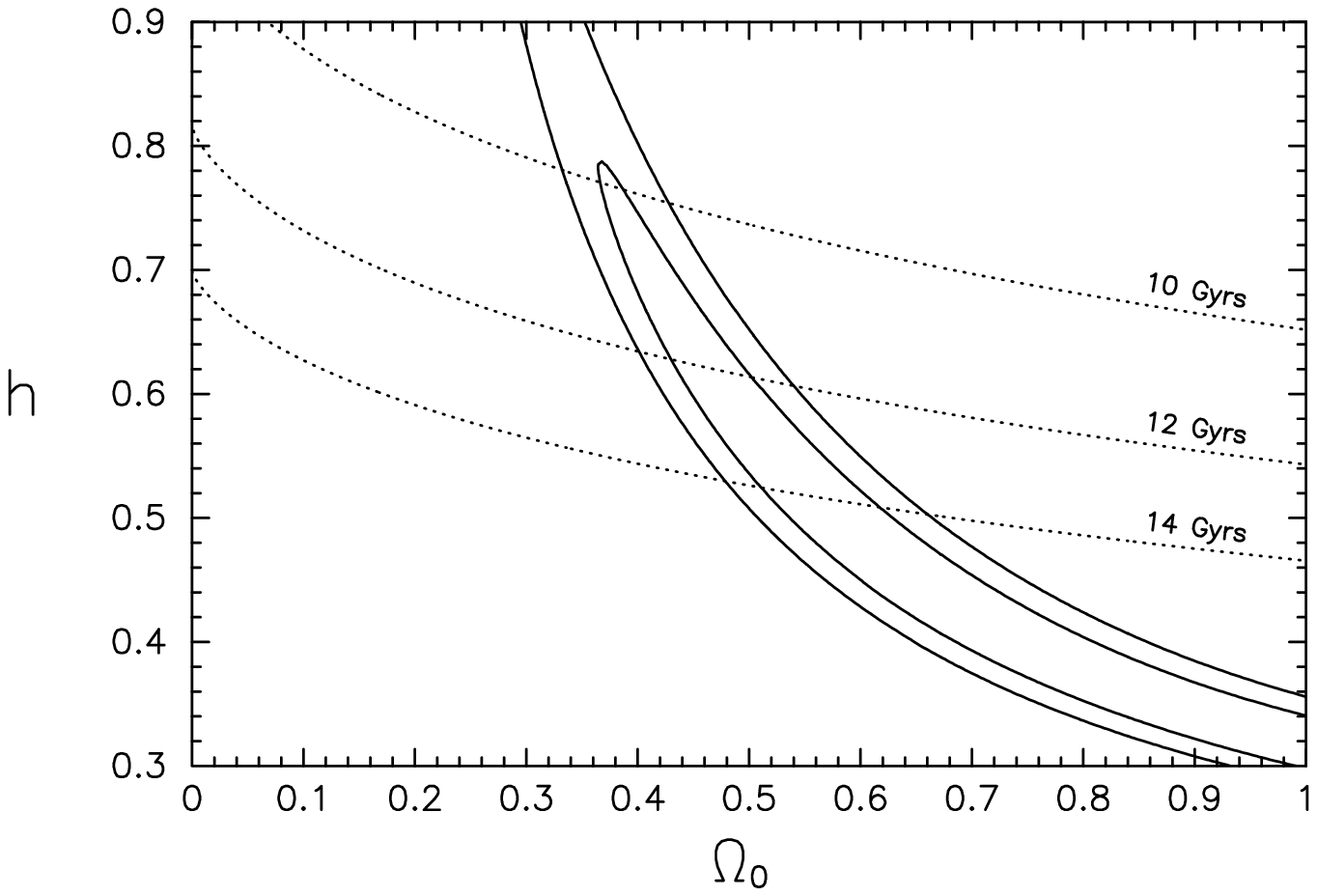}
\caption[figure2]{A chi-squared plot in the $\Omega_0$--$h$ plane. This plot 
includes all the data that we have discussed. In contrast to Fig.~1, the {\it 
COBE} data are included with their uncertainty error bar. The main compromise 
is the condensation of the Peacock \& Dodds data set down to just the shape 
parameter; otherwise its slightly low chi-squared is excessively generous to 
the other data as discussed. The inner contour corresponds to 68 per cent 
confidence, the outer one to 95 per cent confidence. Age contours are also 
shown, but are not included in the chi-squared. The chi-squared plot supports 
the picture formed from Fig.~1, with the drawback of concealing which 
data are primarily responsible for the trends.}
\end{figure*}

At redshift $z=4$, the contribution of the damped Lyman alpha systems to the
density in {\em baryons} is estimated as
\begin{equation}
\Omega_{{\rm DLAS}}=(0.0011 \pm 0.0002) \, h^{-1} \, 
\sqrt{\frac{1+\Omega_0 z}{1+z}} \,,
\end{equation}
where we have conservatively assumed that all the gas in these systems is in 
the neutral state. Remembering that we are taking the dark matter to be cold, 
it is a reasonable hypothesis that the total density of these systems is 
bigger by a factor $\Omega_0/\Omega_{{\rm B}}$, where $\Omega_{{\rm B}} = 
0.0125 h^{-2}$ is the average baryon density given by nucleosynthesis. If $M$ 
is the typical mass of the systems, this implies that the fraction $f(>M,z)$ 
of the total mass that resides in bound objects of mass at least $M$ at 
redshift $z = 4.0$ satisfies 
\begin{equation}
f(>M,z=4.0) > (0.088 \pm 0.024) \, h \, \sqrt{\frac{1+\Omega_0 z}{1+z}}\,,
\end{equation}
where a 20 per cent uncertainty in the baryon fraction has been added in 
quadrature to the observational uncertainty.

Since we want a lower bound on the density perturbation we take the lower end 
of the error bar. Bearing in mind that there is no evidence that damped Lyman 
alpha systems at high redshifts are completely collapsed objects, as we only 
observe their baryonic component which is able to collapse faster through 
radiative cooling (e.g. Katz et al. 1994), we conservatively assume that 
these systems are more akin to collapsing protospheroids (see also Lanzetta 
et al. 1995). In order to reflect this choice we will use $\delta_{{\rm 
c}}=1.5$ in the Press--Schechter calculation, which some numerical studies 
(e.g. Monaco 1995) have shown is associated with the time-scale of 
gravitational collapse of a perturbation along its first two collapsing axes, 
i.e. `filament' formation. Also, in order to be compatible with lower 
redshift observations, the collapsing protospheroids have to be massive 
enough eventually to give rise to rotationally supported discs \cite{LWT}. 
Therefore we take the minimum mass of damped Lyman alpha systems to be 
$10^{10} \, h^{-1} \; {{\rm M}}_{\sun}$ \cite{H95}, which corresponds to a 
circular velocity of about $75 \; {\rm km} \, {\rm s}^{-1}$. Although 
formally the constraint as calculated above is only a 1$\sigma$ lower limit, 
it is almost unchanged by going to 2$\sigma$.

\section{Discussion}

We plot the data that we have discussed in two separate ways. The first is 
direct contouring of the observations in the $\Omega_0$--$h$ plane, in 
Fig.~1. For this figure we have fixed the normalization of the spectrum by 
the {\it COBE} measurement, taking advantage of its small error bar. It turns 
out that the constraint based on the abundance of damped Lyman alpha systems 
is very weak (both in the critical density case and for general $\Omega_0$) 
as compared to other constraints, and so for clarity we do not plot it. All 
other data play some role in constraining the allowed parameters, though the 
shape parameter and cluster abundance allow very similar regions. We use the 
age constraint to cut off the region at the very conservative value of 10 
Gyr.

The second type of plot, Fig.~2, shows chi-squared contours of the data. 
The only difference in input data is that we treat the {\it COBE} data with 
their uncertainty, so that at each point in parameter space the chi-squared 
statistic is that for the optimal normalization. The chi-squared plot has the 
advantage of producing a simple summary of the constraints, but the drawback 
that one cannot tell which of the data are predominant in contributing to 
the constraints. 

Most of the recent literature on structure formation models has concentrated 
attention either on retaining $\Omega_0 = 1$ and making other modifications 
such as introducing a hot dark matter component, or on reducing $\Omega_0$ 
all the way down to 0.2 or 0.3. We notice that the best fits with the new 
{\it COBE} normalization favour rather higher values, the lowest permitted 
being about $\Omega_0 = 0.35$. Further, good fits are available for the whole 
continuum of $\Omega_0$ values above this, for a suitable choice of Hubble 
parameter. Without inputting extra information on the preferred values of 
$h$, the observational data indicate no particular preference for any value 
of $\Omega_0$.

A variety of recent measurements of the Hubble parameter have favoured higher 
values (e.g. Schmidt et al. 1994; Freedman et al. 1994). While we feel that 
the situation has yet to be completely closed, it is interesting to examine 
the reduction in parameter space implied by choosing $h > 0.6$. This still 
allows a fit to all the data (even allowing an age over 12 Gyr), but such a 
constraint requires that $\Omega_0$ falls in a narrow band between 0.35 and 
0.55.

So far in this paper we have assumed that the spectral index of the 
primordial curvature perturbation is $n=1$. Inflation models typically 
predict some degree of `tilt' in the spectrum, so that $n$ is not precisely 
1. The degree of tilt predicted by inflation is highly model-dependent,
ranging from negligible up to a few tenths depending on the inflationary 
model \cite{LL}. There is some preference for tilting to $n < 1$ but 
models also exist that act in the opposite direction. There is some reason to 
believe that tilt is rather more likely in open inflationary models, since 
special physics is being invoked on scales around the curvature scale. We end 
with a short discussion of the effect of tilt.

In the case of critical density, there is a strong desire to remove 
short-scale power from the spectrum, which both tilt and gravitational waves 
are capable of doing. For low densities, the spectrum has already had its 
shape altered by the shifting of matter--radiation equality, and so there is 
less freedom to shift $n$ from the scale-invariant value. In the absence of a 
definite prediction of tilt from an inflationary model, we shall examine the 
two cases $n = 0.9$ and $n = 1.1$. The effect of gravitational waves in an 
open universe has not been successfully quantified yet and we do not include 
them here.

Tilting corresponds to taking $\delta_H^2(\Omega_0)$ to be scale-dependent, 
given by
\begin{equation}
\label{tiltnorm}
\delta_H^2 (k, \Omega_0) = \delta_H^2(\Omega_0) \left( \frac{k}{k_{10}}
	\right)^{n-1} \,,
\end{equation}
where $k_{10}$ is some normalization scale where the normalized spectra at 
different $n$ are assumed to cross (for a given $\Omega_0$). A detailed 
normalization of tilted open models along the lines of G\'{o}rski et al. 
\shortcite{Getal2} has not been provided so we need some improvization to 
normalize the models. We assume that the tenth microwave anisotropy 
multipole, which acts as a pivot point for the {\it COBE} data, is unchanged 
by tilt; hence the notation $k_{10}$ which is the effective scale of the 
tenth multipole. The prefactor $\delta_H^2(\Omega_0)$ is the normalization 
for $n=1$, given by equation (\ref{COBEnorm}). We find the scale $k_{10}$ 
from the `window function' which describes how different scales contribute to 
the tenth multipole; we take $k_{10}$ to be the scale at which the window 
function, calculated using only the Sachs--Wolfe effect as in 
Garc\'{\i}a-Bellido et al. \shortcite{GLLW}, peaks. We find that $k_{10}$ is 
very well fitted by the surprisingly simple relation $k_{10} = 6 \Omega_0 a_0 
H_0$. With this, we can then use equation (\ref{tiltnorm}) in equations 
(\ref{powerspec}) and (\ref{disp}). This improvization should work very well 
until $\Omega_0$ becomes smaller than about 0.3.

All other data remain the same, though we now need a more general expression 
for the shape parameter, which at 95 per cent confidence is \cite{PD}
\begin{equation}
\Gamma = 0.255_{-0.033}^{+0.038} + 0.32 \left( \frac{1}{n} -1 \right) \,.
\end{equation}

The results are shown in Fig.~3. Increasing $n$ has the effect of shifting 
the allowed band to lower $\Omega_0$, and decreasing $n$ shifts it to higher 
$\Omega_0$, roughly in accordance with $\Delta \Omega_0 \simeq - \Delta n/2$. 
Assuming the range $0.9 < n < 1.1$, we therefore find that the width of the 
band in the $\Omega_0$--$h$ plane is increased, so that for $h>0.6$ the 
allowed range is roughly $0.30 < \Omega_0 < 0.60$. 

In conclusion, we have made a thorough investigation of linear theory 
constraints on cold dark matter models in genuinely open universes, on 
the assumption that the spectrum of the primordial curvature perturbation is 
scale-independent. We have also placed these models in their inflationary 
cosmology context. The normalization to {\it COBE} provided by GRSB allows a 
much more precise comparison with observations than has been made previously. 
We have included a treatment of the abundances of both clusters and damped 
Lyman alpha systems; although these have proved constraining for various 
types of model such as mixed dark matter models, they are easy to satisfy 
here. On the whole, the new constraints that we have computed support the 
allowed parameter space from earlier considerations rather than reduce it. We 
conclude that there is a substantial parameter space still viable for these 
models.

\begin{figure}
\centering
\leavevmode\epsfysize=5.8cm \epsfbox{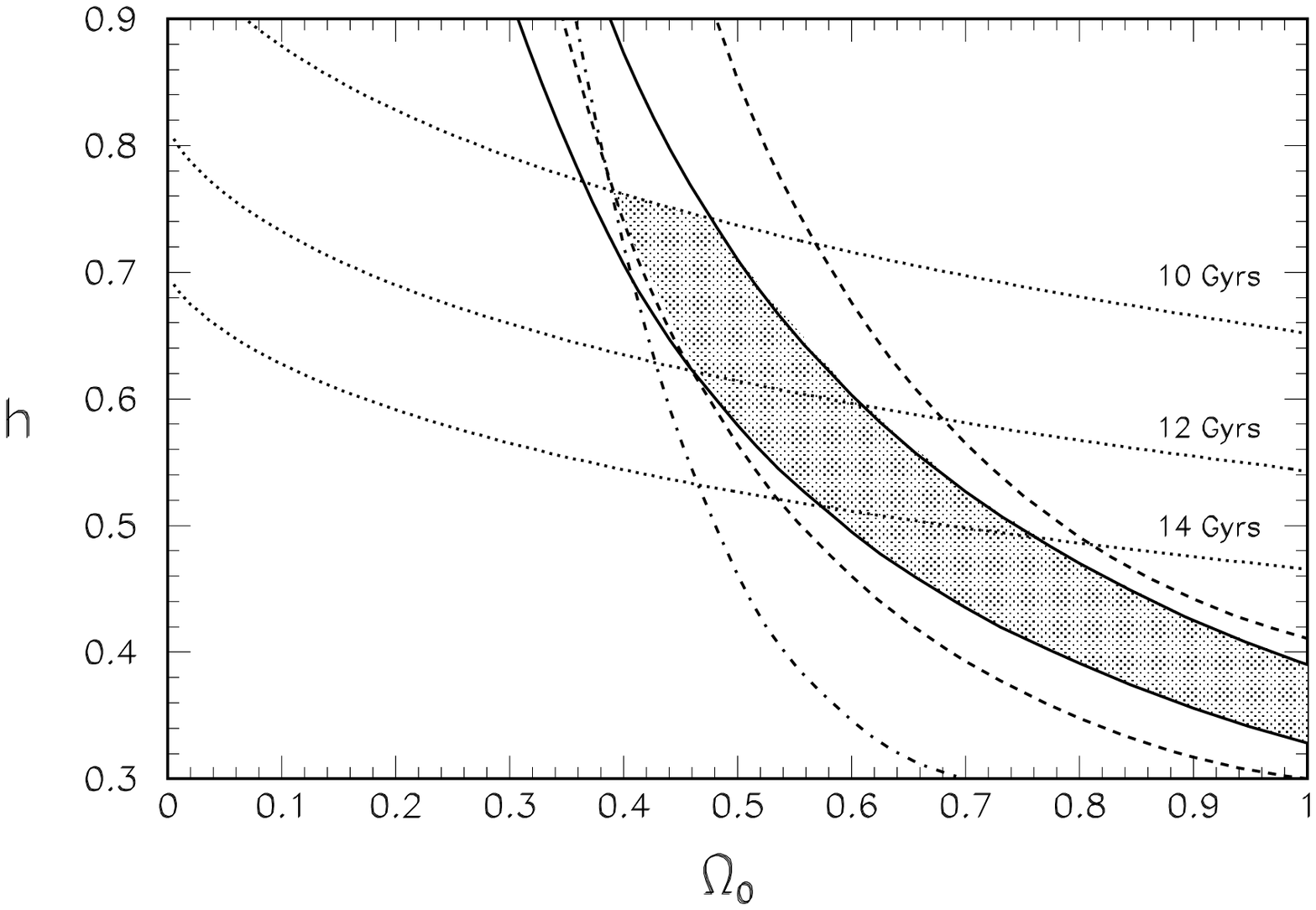}\\
\leavevmode\epsfysize=5.8cm \epsfbox{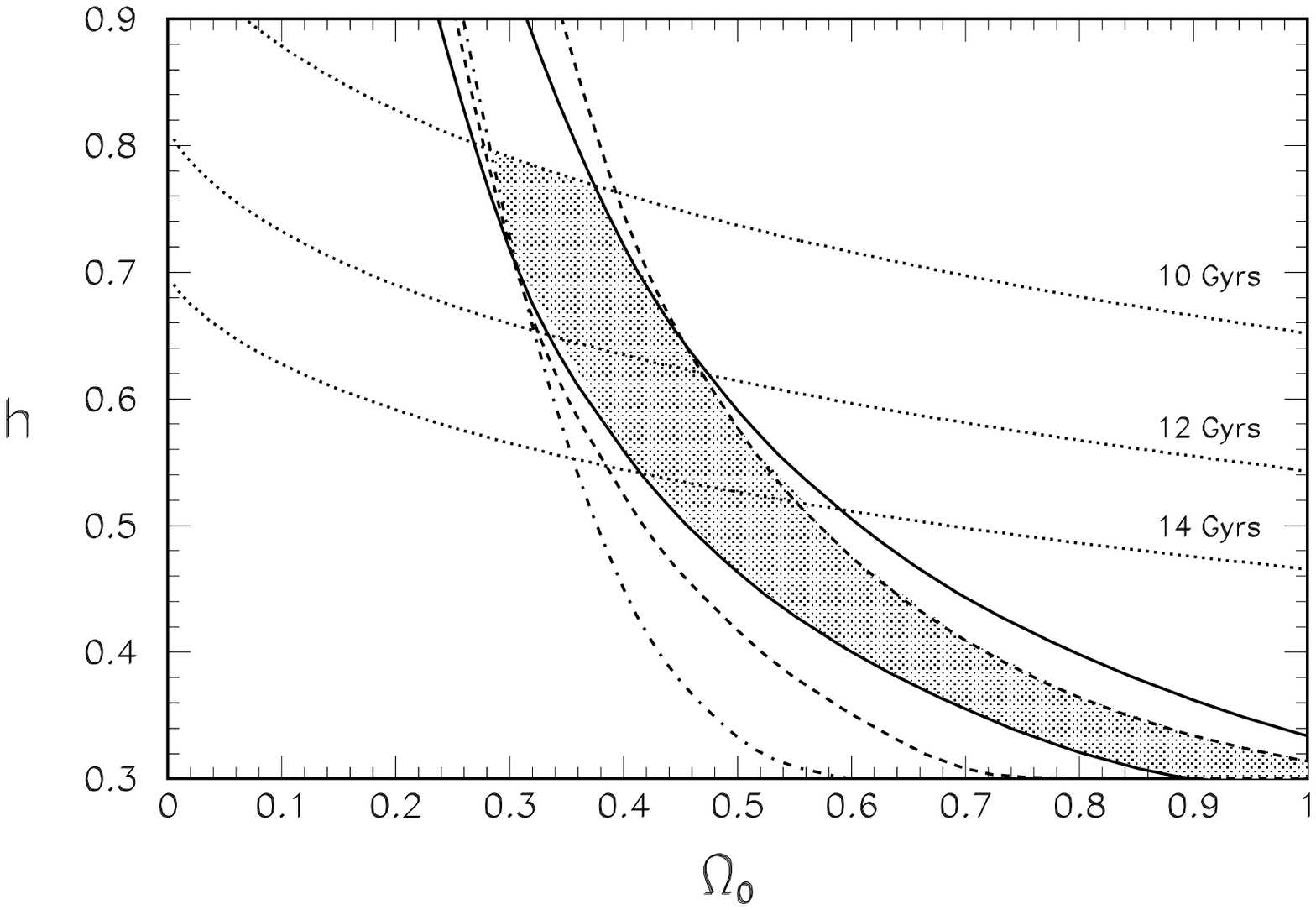}
\caption[figure3]{As Fig.~1, but in the case of a tilted spectrum. The top 
diagram shows $n = 0.9$, and the lower shows $n = 1.1$.}
\end{figure}

\section*{ACKNOWLEDGMENTS}

ARL is supported by the Royal Society, DR by PPARC (UK) and PTPV by the 
PRAXIS XXI program of JNICT (Portugal). ARL and PTPV acknowledge the 
use of the Starlink computer system at the University of Sussex. We thank 
Martin Hendry, Bharat Ratra, Douglas Scott, Naoshi Sugiyama and Martin White 
for comments and discussions, and Juan Garc\'{\i}a-Bellido for help in 
calculating the {\it COBE} normalization for the tilted models. 

\bsp

\begin{thebibliography}{}
\bibitem[\protect\citename{Amendola, Baccigalupi \& Occhionero }1995]
	{bubblef} Amendola L., Baccigalupi C., Occhionero F., 1995,
	Rome pre\-print, astro-ph/9504097
\bibitem[\protect\citename{Bardeen }1980]{Bardeen} Bardeen J. M., 1980, 
	Phys. Rev. D, 22, 1882
\bibitem[\protect\citename{Bardeen et al. }1986]{BBKS} Bardeen J. M., Bond 
	J. R., Kaiser N., Szalay A. S., 1986, ApJ, 304, 15
\bibitem[\protect\citename{Bartelmann \& Loeb }1995]{bartelmann} 
	Bartelmann M., Loeb A., 1995, CfA preprint, astro-ph/9505078
\bibitem[\protect\citename{Bennett et al. }1994]{B94} Bennett C. L. et al.,
	1994, ApJ, 436, 423
\bibitem[\protect\citename{Bernardeau }1994]{Ber} Bernardeau F.,
	1994, ApJ, 427, 51
\bibitem[\protect\citename{Bertschinger \& Dekel }1989]{BD89} Bertschinger
	E., Dekel A., 1989, ApJ, 336, L5
\bibitem[\protect\citename{Bruni \& Lyth }1994]{brly} Bruni M., Lyth D. H.,
	1994, Phys. Lett., B323, 118
\bibitem[\protect\citename{Bucher \& Turok }1995]{bt} Bucher M., Turok N.,
	1995, Princeton preprint, hep-ph/9503393
\bibitem[\protect\citename{Bucher et al. }1995]{BGT} Bucher M., Goldhaber A. 
	S., Turok N., 1995, Phys. Rev. D, 52, 3314 
\bibitem[\protect\citename{Carroll, Press \& Turner }1992]{CPT} Carroll 
	S. M., Press W. H., Turner E. L., 1992, ARA\&A, 30, 499
\bibitem[\protect\citename{Colafrancesco \& Vittorio }1994]{CV} 
	Colafrancesco S., Vittorio N., 1994, ApJ, 422, 443
\bibitem[\protect\citename{Coleman \& de Luccia }1980]{bubblea} 
	Coleman S., de Luccia F., 1980, Phys. Rev. D, 21,  3305
\bibitem[\protect\citename{Coles \& Ellis }1994]{ce} Coles P., Ellis G. 
	F. R., 1994, Nat, 370, 609
\bibitem[\protect\citename{Copi, Schramm \& Turner }1995]{CST} Copi C. J., 
	Schramm D. N., Turner M. S., 1995, Science, 267, 192
\bibitem[\protect\citename{Dekel }1994]{Dekel} Dekel A., 1994, ARA\&A, 32, 
	371
\bibitem[\protect\citename{Demarque et al. }1991]{DDS} Demarque P.,
	Deliyannis C. P., Sarajedini A., 1991, in Shanks T., Banday A. J.,
	Ellis R. S., Frenk C. S., Wolfendale A. W., eds, Observational Tests
	of Cosmological Inflation, Kluwer, Dordrecht, p.~111
\bibitem[\protect\citename{Efstathiou, Sutherland \& Maddox }1990]{ESM}
	Efstathiou G., Sutherland W. J., Maddox S. J., 1990, Nat, 348, 705
\bibitem[\protect\citename{Evrard }1989]{E89} Evrard A. E., 1989, ApJ, 
	341, L71
\bibitem[\protect\citename{Evrard }1990]{E} Evrard A. E., 1990, ApJ, 363, 
	349
\bibitem[\protect\citename{Freedman et al. }1994]{H0} Freedman W. L.
	et al., 1994, Nat, 371, 757
\bibitem[\protect\citename{Garc\'{\i}a-Bellido et al. }1995]{GLLW}
	Garc\'{\i}a-Bellido J., Liddle A. R., Lyth D. H., Wands D., 
	1995, to appear, Phys. Rev. D, astro-ph/9508003
\bibitem[\protect\citename{G\'orski et al. }1994]{Getal} G\'orski K. M.
	et al., 1994, ApJ, 430, L89
\bibitem[\protect\citename{G\'orski et al. }1995]{Getal2} G\'orski K. M.,
	Ratra B., Sugiyama N., Banday A. J., 1995, ApJ, 444, L65 [GRSB]
\bibitem[\protect\citename{Gott }1982]{bubbleb} Gott J. R., 1982, Nature,
	295, 304
\bibitem[\protect\citename{Gott \& Statler }1984]{bubbled}
	Gott J. R., Statler T. S., 1984, Phys. Lett., B136, 157
\bibitem[\protect\citename{Guth \& Weinberg }1983]{bubblec} 
	Guth A. H., Weinberg E. J., 1983, Nucl. Phys., B212, 321
\bibitem[\protect\citename{Haehnelt }1995]{H95} Haehnelt M. G., 1995, 
	MNRAS, 273, 249
\bibitem[\protect\citename{Hanami }1993]{H} Hanami H., 1993, ApJ, 415, 42
\bibitem[\protect\citename{Henry \& Arnaud }1991]{HA} Henry J. P., 
	Arnaud K. A., 1991, ApJ, 372, 410
\bibitem[\protect\citename{Kamionkowski \& Spergel }1994]{KamS} Kamionkowski
	M., Spergel D. N., 1994, ApJ, 432, 7
\bibitem[\protect\citename{Kamionkowski et al. }1994]{KRSS} Kamionkowski
	M., Ratra B., Spergel D. N., Sugiyama N., 1994, ApJ, 434, L1 
\bibitem[\protect\citename{Katz et al. }1994]{KQBG} Katz N., Quinn T., 
	Bertschinger E., Gelb J. M., 1994, MNRAS, 270, L71
\bibitem[\protect\citename{Kauffmann \& Charlot }1994]{KC} Kauffmann G.,
	Charlot S., 1994, ApJ, 430, L97
\bibitem[\protect\citename{Klypin et al. }1995]{KDLAS} Klypin A., Borgani S.,
	Holtzman J., Primack J., 1995, ApJ, 444, 1
\bibitem[\protect\citename{Kofman, Gnedin \& Bahcall }1993]{KGB} Kofman L.
	A., Gnedin N. Y., Bahcall N. A., 1993, ApJ, 413, 1
\bibitem[\protect\citename{Lacey \& Cole }1993]{LC93} Lacey C., Cole S.,
	1993, MNRAS, 262, 627
\bibitem[\protect\citename{Lacey \& Cole }1994]{LC} Lacey C., Cole S., 1994,
	MNRAS, 271, 676
\bibitem[\protect\citename{Lanzetta et al. }1995]{LWT} Lanzetta K. M., 
	Wolfe A. M., Turnshek D. A., 1995, ApJ, 440, 435
\bibitem[\protect\citename{Liddle \& Lyth }1993]{LL} Liddle A. R., Lyth D. 
	H., 1993, Phys. Rep., 231, 1
\bibitem[\protect\citename{Liddle \& Lyth }1995]{LL95} Liddle A. R., Lyth D. 
	H., 1995, MNRAS, 273, 1177
\bibitem[\protect\citename{Lilje }1992]{lilje} Lilje P. B., 1992, ApJ,
	386, L33
\bibitem[\protect\citename{Linde }1995]{bubblee} Linde A., 1995, Phys.
	Lett., B351, 99
\bibitem[\protect\citename{Linde \& Mezhlumian }1995]{LinMez} Linde A., 
	Mezhlumian A., 1995, Stanford preprint, astro-ph/9506017
\bibitem[\protect\citename{Lyth \& Stewart }1990a]{lyst} Lyth D. H., Stewart
	E. D., 1990a, Phys. Lett., B252, 336
\bibitem[\protect\citename{Lyth \& Stewart }1990b]{lystb} Lyth D. H., Stewart
	E. D., 1990b, ApJ, 361, 343
\bibitem[\protect\citename{Lyth \& Woszczyna }1995]{LW} Lyth D. H., Woszczyna 
	A., 1995, Phys. Rev. D, 52, 3338
\bibitem[\protect\citename{Ma \& Bertschinger }1994]{MB} Ma C.-P., 
	Bertschinger E., 1994, ApJ, 434, L5
\bibitem[\protect\citename{Metzler \& Evrard }1994]{ME} Metzler C. A., 
	Evrard A. E., 1994, ApJ, 437, 564
\bibitem[\protect\citename{Mo \& Miralda-Escud\'{e} }1994]{MM} Mo H. J., 
	Miralda-Escud\'{e} J., 1994, ApJ, 430, L25
\bibitem[\protect\citename{Monaco }1995]{Mo} Monaco P., 1995, 
	ApJ, 447, 23
\bibitem[\protect\citename{Navarro et al. }1995]{NFW} Navarro J. 
	F., Frenk C. S., White S. D. M., 1995, MNRAS, 275, 720
\bibitem[\protect\citename{Peacock \& Dodds }1994]{PD} Peacock J. A., Dodds 
	S. J., 1994, MNRAS, 267, 1020
\bibitem[\protect\citename{Press \& Schechter }1974]{PS} Press W. H.,
	Schechter P., 1974, ApJ, 187, 452
\bibitem[\protect\citename{Primack }1995]{Prim} Primack J. R., 1995, Santa
	Cruz preprint, astro-ph/9503020
\bibitem[\protect\citename{Ratra \& Peebles }1994]{RPa} Ratra B., Peebles
	P. J. E., 1994, ApJ, 432, L5
\bibitem[\protect\citename{Ratra \& Peebles }1995]{RPb} Ratra B., Peebles
	P. J. E., 1995, Phys. Rev. D, 52, 1837
\bibitem[\protect\citename{Sasaki et al. }1993a]{one} Sasaki M., Tanaka T.,
	Yamamoto K., Yokoyama J., 1993a, Phys. Lett., B317, 510
\bibitem[\protect\citename{Sasaki et al. }1993b]{two} Sasaki M., Tanaka T.,
	Yamamoto K., Yokoyama J., 1993b, Prog. Theor. Phys., 90, 1019
\bibitem[\protect\citename{Sasaki et al. }1995]{five} 
	Sasaki M., Tanaka T., Yamamoto K., 1995, Phys. Rev. D, 51, 2979
\bibitem[\protect\citename{Sasaki }1994]{Sa} Sasaki S., 1994, PASJ, 46, 427
\bibitem[\protect\citename{Schmidt et al. }1994]{H02} Schmidt B. P.,
	Kirschner R. P., Eastman R. G., Phillips M. M., Suntzeff N. B., 
	Hamuy M., Maza J., Avil\'{e}s R., 1994, ApJ, 432, 42
\bibitem[\protect\citename{Smoot et al. }1992]{COBE} 
	Smoot G. F. et al., 1992, ApJ, 396, L1
\bibitem[\protect\citename{Stockton et al. }1995]{SKR} Stockton A., 
	Kellogg M., Ridgway S. E., 1995, ApJ, 443, L69
\bibitem[\protect\citename{Storrie-Lombardi et al. }1995]{storrie} 
	Storrie-Lombardi L. J., McMahon R. G., Irwin M. J., Hazard C., 1995,
	Cambridge preprint, astro-ph/9503089
\bibitem[\protect\citename{Sugiyama }1994]{SUG94} Sugiyama N., Berkeley
	preprint, astro-ph/9412025
\bibitem[\protect\citename{Sugiyama \& Silk }1994]{SSilk} Sugiyama N., 
	Silk J., 1994, Phys. Rev. Lett., 73, 509
\bibitem[\protect\citename{Tanaka \& Sasaki }1994]{three} 
	Tanaka T., Sasaki M., 1994, Phys. Rev. D, 50, 6444
\bibitem[\protect\citename{Viana \& Liddle }1995]{VL} Viana P. T. P., Liddle
	A. R., 1995, Sussex preprint
\bibitem[\protect\citename{Walker et al. }1991]{NUCL} Walker T., Steigman G.,
	Schramm D. N., Olive K. A., Kang H.-S., 1991, ApJ, 376, 51
\bibitem[\protect\citename{White \& Bunn }1995]{WB} White M., Bunn E. F.,
	1995, ApJ, 450, 477
\bibitem[\protect\citename{White et al. }1993a]{WEF} White S.
	D. M., Efstathiou G., Frenk C. S., 1993a, MNRAS, 262, 1023
\bibitem[\protect\citename{White et al. }1993b]{WNEF} White S. D. M.,
	Navarro J. F., Evrard A. E., Frenk C. S., 1993b, Nat, 366, 429
\bibitem[\protect\citename{Wright et al. }1994]{W94} Wright E. L. et al.,
	1994, ApJ, 420, 1
\bibitem[\protect\citename{Yamamoto \& Bunn }1995]{YB} Yamamoto K., Bunn
	E. F., 1995, Berkeley preprint, astro-ph/9508090
\bibitem[\protect\citename{Yamamoto et al. }1995a]{sasaki} 
	Yamamoto K., Sasaki M., Tanaka T., 1995a, Kyoto preprint,
	astro-ph/9501109
\bibitem[\protect\citename{Yamamoto et al. }1995b]{four} 
	Yamamoto K., Tanaka T., Sasaki M., 1995b, Phys. Rev. D, 51, 2968
\end{thebibliography}
\end{document}